\documentclass[sigconf]{acmart}

\AtBeginDocument{%
  }

\usepackage{subcaption}
\usepackage{enumitem}

\copyrightyear{2024}
\acmYear{2024}
\setcopyright{acmlicensed}
\acmConference[RecSys '24]{18th ACM Conference on Recommender Systems}{October 14--18, 2024}{Bari, Italy}
\acmBooktitle{18th ACM Conference on Recommender Systems (RecSys '24), October 14--18, 2024, Bari, Italy}
\acmDOI{10.1145/3640457.3688135}
\acmISBN{979-8-4007-0505-2/24/10}

\usepackage{multirow}

\begin{document}

\title{LARR: Large Language Model Aided Real-time Scene Recommendation with Semantic Understanding}


\author{Zhizhong Wan}
\email{wanzhizhong@meituan.com}
\orcid{0009-0005-8624-9249}
\affiliation{
\institution{Meituan}
 \city{Beijing}
  \country{China}
}
\author{Bin Yin}
\email{yinbin05@meituan.com}
\orcid{0009-0007-3228-2670}
\affiliation{
\institution{Meituan}
 \city{Beijing}
  \country{China}
}
\author{Junjie Xie}
\email{xiejunjie02@meituan.com}
\orcid{0009-0009-2793-0092}
\affiliation{
\institution{Meituan}
 \city{Beijing}
  \country{China}
}
\author{Fei Jiang}
\email{jiangfei05@meituan.com}
\orcid{0000-0002-7019-140X}
\affiliation{
\institution{Meituan}
 \city{Beijing}
  \country{China}
}
\author{Xiang Li}
\email{lixiang245@meituan.com}
\orcid{0000-0003-2834-8765}
\affiliation{
\institution{Meituan}
 \city{Beijing}
  \country{China}
}
\author{Wei Lin}
\email{linwei31@meituan.com}
\orcid{0000-0003-2851-820X}
\affiliation{
\institution{Meituan}
 \city{Beijing}
  \country{China}
}


\begin{abstract}
Click-Through Rate (CTR) prediction is crucial for Recommendation System(RS), aiming to provide personalized recommendation services for users in many aspects such as food delivery, e-commerce and so on. However, traditional RS relies on collaborative signals, which lacks semantic understanding to real-time scenes. We also noticed that a major challenge in utilizing Large Language Models (LLMs) for practical recommendation purposes is their efficiency in dealing with long text input. To break through the problems above, we propose Large Language Model Aided Real-time Scene Recommendation(LARR), adopt LLMs for semantic understanding, utilizing real-time scene information in RS without requiring LLM to process the entire real-time scene text directly, thereby enhancing the efficiency of LLM-based CTR modeling. Specifically, recommendation domain-specific knowledge is injected into LLM and then RS employs an aggregation encoder to build real-time scene information from separate LLM's outputs. Firstly, a LLM is continual pretrained on corpus built from recommendation data with the aid of special tokens. Subsequently, the LLM is fine-tuned via contrastive learning on three kinds of sample construction strategies. Through this step, LLM is transformed into a text embedding model. Finally, LLM's separate outputs for different scene features are aggregated by an encoder, aligning to collaborative signals in RS, enhancing the performance of recommendation model.
\end{abstract}



\received{29 April 2024}
\received[revised]{22 July 2024}
\received[accepted]{19 August 2024}

\maketitle

\section{Introduction}
As an important task in field of recommendation systems (RS), Click-Through Rate (CTR) prediction aims to forecast whether a user will click on a certain product, content, or advertisement. In the domain of RS, models need to analyze data such as user characteristics and item features to predict the probability of a user clicking on a specific objective, thereby helping to optimize the effectiveness of recommendation strategies. 

In the context of food delivery service, the volume of data that needs to be processed each day reaches hundreds of millions. Unlike typical e-commerce recommendations or content recommendations, food delivery recommendations place a stronger emphasis on analyzing real-time scenes such as geographical location, mealtime, weather, etc. Among them, there exist strong correlations and rich semantics. Consider a specific scene: a user traveling to another city arrives late at night while many restaurants are already closed, and it is raining heavily with low temperatures. Due to the user's unfamiliarity with the surrounding environment and the poor weather conditions, there is a high probability that he will order a takeout from a nearby restaurant. What's more, he's more likely to order some hot food or local specialties. 

Points of Interest (POI) is a concept we define based on food delivery recommendations, aimed at abstracting the focus of user interest. It can be understood as a collection of POIs, environment, and geographic locations. Traditional recommendation models use feature crossing to handle POIs in real-time scene data. For correlated and important features, common feature crossing methods include manual explicit crossing followed by logistic regression \cite{cheng2016wide}, explicit low-order crossing \cite{cheng2016wide}, or directly using deep learning for automatic high-order crossing \cite{guodeepfm, huang2019fibinet, tang2020progressive}. However, regardless of which method is used, the essence of feature crossing is based on the co-occurrence probability of strongly correlated features to determine whether a user will click, lacking an understanding of the POI in the scene's semantic information. 

Large Language Models (LLMs) possess extensive semantic knowledge and excellent reasoning capabilities \cite{hua2023tutorial,he2023large}, which can effectively understand the semantic information in real-time scenes. However, in the field of RS, LLMs still face some difficulties. In particular, LLMs in their original form may not possess in-depth knowledge specific to the recommendation domain, which could affect their comprehensive understanding of POI and user information. At the same time, LLMs typically do not directly handle the cross-feature signals present in traditional recommendation models, which may limit their ability to accurately capture and understand user needs.

\begin{figure}[h]
  \centering
  \includegraphics[width=\linewidth]{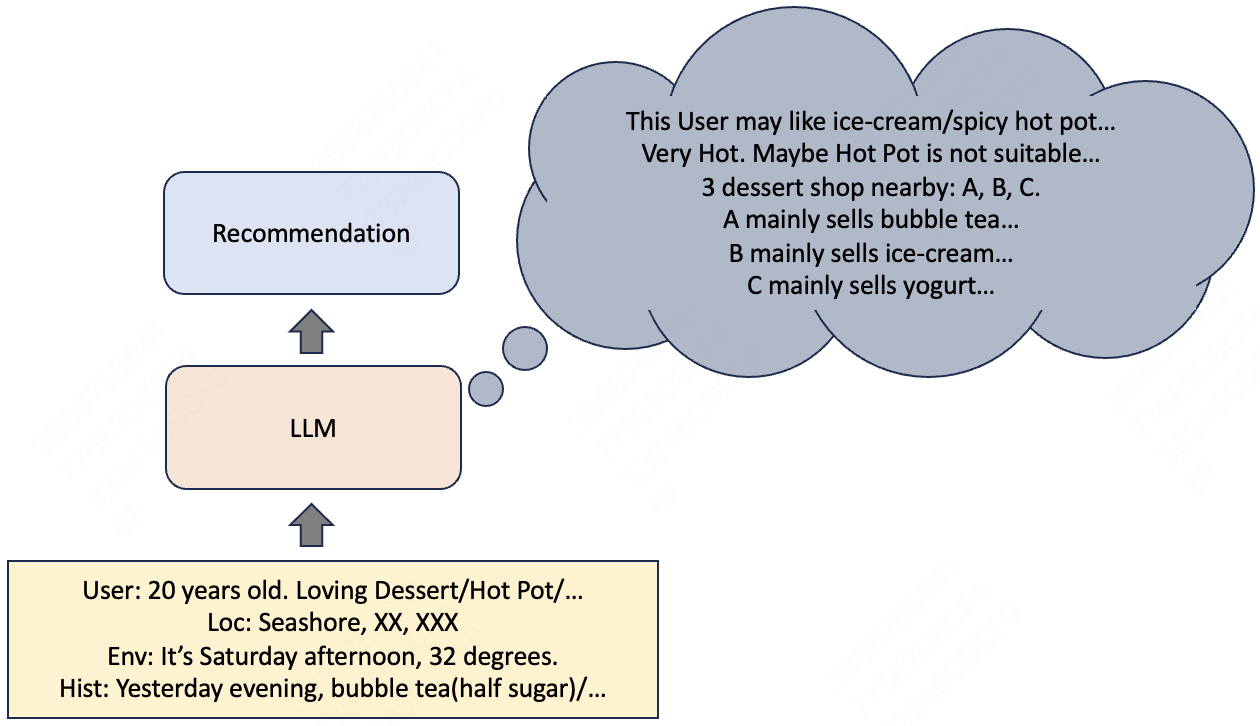}
  \caption{How LLM understands the real-time scenes and help Recommendation System work.}
  \Description{intuitive fig 1}
  \label{fig:1}
\end{figure}

The unique recommendation scene of food delivery presents us with two core challenges:
\begin{itemize}[leftmargin=*]
\item Typical RS treats each food delivery POI as mapped unique id token as inputs of the recommendation model to predict user behavior. However, this approach treats all POI equally without considering semantic information of POIs, such as the menu similarity of food delivery, relevance of restaurant's main business direction to the scene and so on. For instance, pizza shops with different names would be treated as two unrelated POIs in a normal RS, neglecting the fact that both of them primarily sell pizza. During summer, people tend to order something cold rather than taste spicy and hot food. 

\item Industrial RS needs to serve for millions of users, unacceptable time consuming of LLM's inference stays a unsolved problem in the exploring of combination with RS and LLM. Real-time scenes are composed of many individual scene features, with combinations of different scene features giving rise to an exponential number of different real-time scenes. To understand real-time scenes, LLMs need to process combinations of scene feature texts during serving, which seems that LLMs have to be involved in inference, leading to service latency. How to efficiently utilize LLMs to handle real-time scenes is an urgent problem to be solved.

\end{itemize}
To address the two problems mentioned above, inspired by recent work in the LLM field \cite{neelakantan2022text,li2023ctrl} and recommendation system domain \cite{geng2022recommendation,cui2022m6}, we propose a three-stage model LARR (Large Language Model Aided Real-time Scene Recommendation with Semantic Understanding). Figure ~\ref{fig:1} uses an intuitive example to show how LARR works. 
The LLM, infused with recommendation domain knowledge, processes various scene information based on its knowledge, such as user profiles, geographic locations, weather, etc. The LLM contemplates the input information according to its existing knowledge, assisting the recommendation system in providing recommendation services. In the example of Figure ~\ref{fig:1}, LLM receives input information including user profiles, geographic locations, environment, user history and so on. Because LLM is injected with recommendation domain knowledge in advance, LLM knows what style of food the user may like, which kind of food is more popular in current environment, where to find tasty shops nearby. From the user profile (likes desserts/hotpot), LLM knows ice-cream or spicy hot pot is ok; furthermore, since it is very hot, hot pot is obviously not quite suitable, ice cream or a cold drink would be a better choice. There are three stores A, B, and C selling desserts around the input geographic location, and their takeout information is... Like this, step by step, the LLM finally summarizes useful information and contributes to RS in recommending takeout for the user.

Our core contributions are as follows:
\begin{itemize}[leftmargin=*]
\item Based on food delivery domain-specific recommendation data, We employ a LLM to deal with the semantic problem of how to correctly understand the POI in food delivery real-time scene. We continue pretraining and fine-tuning the LLM, infusing them with knowledge of recommendation field, promoting the LLMs' understanding of food delivery real-time scenes. 

\item We propose a novel industrial-friendly recommendation framework, LARR, which can effectively integrate the semantic information of food delivery real-time scenes without LLM's participation in inference. We use a bidirectional encoder to aggregate the separate real-time scene embedding produced by LLM to extract the semantic information of real-time scenes, since the semantic associations between various scene features have already been implicitly encoded into the LLM's output vectors.

\item We conducted extensive offline and online experiments on the food delivery dataset from Meituan Waimai, validating that our method can fully understand the semantic information of real-time scenes and effectively integrate multimodal features, enhancing the recommendation results to align the semantic information understood by large language models about the food delivery real-time scene with the recommendation models based on collaborative signals.
\end{itemize}

\section{Related Work}
\subsection{Large Language Model}
Substantial work \cite{devlin2018bert,zhang2019ernie,lewis2019bart} has shown that Pre-trained models (PTMs) on a large corpus can learn universal language representations, which are beneficial for downstream NLP tasks and can avoid training a new model from scratch \cite{qiu2017pre}. Large language models (LLMs) mainly refer to transformer-based \cite{vaswani2017attention} neural language models that contain tens to hundreds of billions of parameters \cite{zhao2023survey}. Those language models are pre-trained on massive text data, such as PaLM \cite{chowdhery2023palm}, LLaMA \cite{touvron2023llama}, and GPT-4 \cite{achiam2023gpt}. Compared to PLMs, LLMs are not only significantly larger in model size but also demonstrate superior language understanding and generation capabilities, more importantly, they exhibit emergent abilities \cite{wei2022emergent} that are absent in smaller-scale language models. 

LLMs have already drawn a lot of attention due to the strong performance on a wide range of natural language tasks, many researchers in other fileds try to combine LLM with domain-specific work to promote their progress and has achieved success \cite{li2022blip,li2023blip,xi2023rise}.

\subsection{LLM in Recommend System}
Traditional recommendation models are designed to leverage a huge amount of ID tokens to train, which is an ID paradigm, with the majority of parameters concentrated in the embedding layer \cite{lian2018xdeepfm,tang2020progressive}. In contrast, LLMs would use a tokenizer to segment the text into vocab tokens to reduce the size of the vocabulary at the very first of input, which is a tokenization paradigm, with the bulk of parameters being concentrated in the network itself. Recommendation models excel in memorization, while LLMs demonstrate superior capabilities in logical reasoning and generalization. A 6-billion parameter LLM such as ChatGLM-3 \cite{du2022glm} has a vocabulary less than 65,000 tokens, a scale that is significantly smaller than the number of ID tokens in a recommendation model of the same scale \cite{ma2018modeling}. It's apparent that simple integration of recommendation systems with LLMs is infeasible. If ID tokens are directly used as input, tokenization may establish connections between unrelated IDs (e.g., 'id\_499' tokenized into 'id', '\_', '4', '9', '9', creating overlapping embeddings with IDs containing the digits 4 and 9, which does not meet our expectations). Alternatively, regard the ID tokens as special tokens of LLM's vocabulary, that is, without tokenization on ID input, would face gap problem between ID tokens representing user/item collaborative information and pre-trained vocab tokens holding content semantics information \cite{bao2023tallrec}. Moreover, the typically vast number of ID tokens could dilute the LLM's own vocabulary, inject noise into the vocab tokens, and result in poor learning of the ID tokens.

Despite the challenges mentioned above associated with applying language models to the field of recommendation, researchers continue to dedicate efforts to applying language models to Recommender Systems (RS), due to the astonishing capabilities demonstrated by LLMs \cite{wei2024llmrec,li2023e4srec}.
The researchers of P5 \cite{geng2022recommendation} proposed a unified text-to-text paradigm recommendation model based on T5 \cite{raffel2020exploring}, hoping to handle rating prediction task, sequential recommendation task and more downstream tasks by zero-shot or few-shots. An embedding method called  whole-word embedding whose design inspiration is very similar to position embedding is introduced to address ID-related gap problem mentioned above. 
M6-rec \cite{cui2022m6} converts all recommend downstream tasks into language understanding or generation tasks by representing user behavior data and candidates data if necessary as natural language plain texts for fine-tuning based on M6 \cite{lin2021m6}. 
CLLM4Rec \cite{zhu2023collaborative} propose a novel soft/hard prompting strategy, mutually-regularized pre-training two LLMs and two set of id tokens on two corpora to facilitate language modeling on RS-specific corpora with heterogeneous user/item collaborative tokens and content tokens.

As an unsupervised method, Contrastive Learning (CL) assumes some observed pairs of text that are more semantically similar than randomly sampled text. By maximazing their mutual information, neural network could learn useful embeddings for downstream tasks \cite{arora2019theoretical,zhang2023contrastive}. OpenAI has attempted to convert pre-trained language models into vector models for text and code using contrastive learning, achieving notable results \cite{neelakantan2022text}. 
RLMRec \cite{ren2023representation} has proven that contrastive learning between the semantic embeddings from LLMs and the collaborative  embeddings from recommender systems could capture their shared information and alleviate the noise information. 
ControlRec \cite{qiu2023controlrec} uses contrastive learning for heterogeneous feature matching to align the ID representations with the natural language in the semantic space. 
Li proposed a two-stage model, CTRL \cite{li2023ctrl}. In the first stage, an LLM is used to encode textual data and a lightweight collaborative model is used to encode tabular data, cross-modal contrastive learning is employed to fine-grained align knowledge of the two modalities. In the second stage, the lightweight collaborative model is fine-tuned on downstream tasks. During inference, CTRL only deploys the collaborative model, with the LLM not being involved in computation, making it an industrial-friendly model architecture.

\section{Methodology}
\begin{figure}[h]
  \centering
  \includegraphics[width=\linewidth]{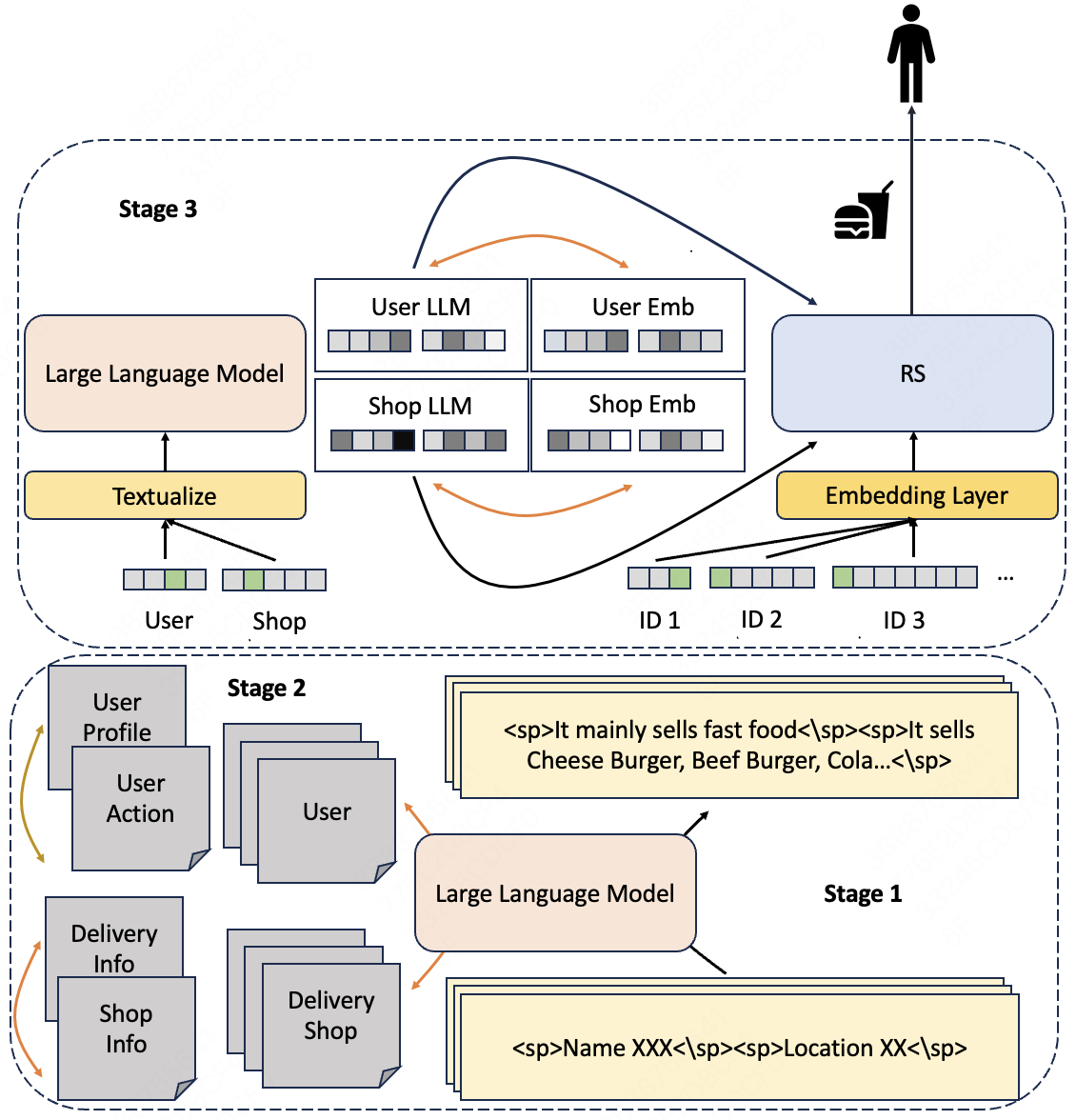}
  \caption{Model overview including 3 stages. In stage 1, The LLM undergoes a continual pretraining task on shop-related corpus; In stage 2, LLM is fine-tuned and transformed into a text embedding model via contrastive learning on 3 types of positive samples; In stage 3, alignment is applying on LLM's semantic embedding and RS's collaborative embedding for enhancing the performance of the recommendation results.}
  \Description{Model overview fig 2}
  \label{fig:2}
\end{figure}

In this section, we will then introduce the framework of LARR, which comprises three stages, with the overall structure illustrated in Fig ~\ref{fig:2}. 
The first stage is the continual pretraining stage, where we construct natural language texts corpus from the Meituan Waimai dataset. Then we continue pretraining the LLM that has been pre-trained on general corpora based on corpus to inject the domain-specific knowledge. 
In the second stage, we additionally add billions of user profile and user history behavior data into the corpus, which is proven to be useful \cite{feng2024context}, constructing 3 kinds of contrastive learning positive and negative samples, transforming the LLM into a text embedding model, enabling the LLM to fully understand the semantics of real-time scenes. 
The final third stage is the multi-modal alignment stage, where contrastive learning is used to maximize the mutual information between semantic embeddings and collaborative embeddings, aligning the takeout scene semantic information understood by the LLM with the collaborative signals extracted by the fine-tuning model from ID tokens, cross features, and statistical features. The aligned semantic information will enhance performance of recommendation system.
\subsection{Continual Pretraining Task Design}
To facilitate LLM's understanding in food delivery domain-specific knowledge from the natural language perspective, we first construct a corpus using datasets relevant to all POIs. Following this, we perform continual pretraining task on the LLM which had already been pre-trained on generic corpora, using this corpus to enhance its understanding of domain-specific knowledge.

How to use id tokens in LLM efficiently remains an unsolved problem for a long-time. In industrial scene, encoding the ids of millions of different POIs as whole words is impractical due to the high time and memory costs involved. Moreover, simply treating POI id tokens as indivisible inputs to the LLM brings a semantic and collaborative signal gap, as mentioned in section 2.2; segmenting and encoding ids could also introduce unintended connections between id tokens. Consequently, we abandoned the use of id token as LLM inputs and instead used unique natural language description texts to represent id features. We noted that a POI can be uniquely identified by its its name and geographical location, which is equivalent to an id feature. The advantage of using name and geographical location to represent a POI is apparently: natural language input makes it easier for LLM to understand. It retains the uniqueness of id inputs while avoiding the semantic meaninglessness of id inputs.

Specifically, assuming that there are $N_{p}$ POIs, we merge name, geographical location, introduction, and statistical information to form a description $D_i$ for each POI $POI_i$. 
\begin{equation}
D = \{ D_i \mid i \in \mathbb{N}, 0 \leq i \leq N_{p} \}
\end{equation}

The description $D_i$ is cut into $n_k$ different slices and we named each slice a keyword, such as name, location, tag name... The $i^{th}$ restaurant's description $D_i$ could be formulated as a text set $t_i$:
\begin{equation}
t_i = \{ {t_i}^k \mid k \in \mathbb{N}, 0 \leq k \leq n_k \}
\end{equation}
\begin{equation}
D_i =  [{t_i}^0, {t_i}^1,...{t_i}^{n_k}]
\end{equation}
where $[...]$ is concatenation between different text and ${t_i}^k$ is a discrete text feature, ${t_i}$ is the aggregation of a bunch of discrete text features.

Through the descriptive text, the LLM could capture the similarity among different POIs. Since the language model was pre-trained on general corpora, thus there are some gap between the pretraining and continual pretraining corpora. To address this problem, we introduce some special tokens to assist the LLM in better understanding the content during the continual pretraining stage. To be specific, we use a set of special token $sp$ for $n_k$ keywords in description to wrap key information. 
\begin{equation}
sp_{p} = \{ (sp^{bos}_k, sp^{eos}_k) \mid k \in \mathbb{N}, 0 \leq k \leq n_k \}
\end{equation}

For example, <POI\_name> and </POI\_name> would be added at the very first and the very last of the $0^{th}$ keyword POI name respectively, the former special token is $sp^{bos}_0$ indicating that the following information pertains to the restaurant's name while the trailing one is is $sp^{eos}_0$, signaling the end of the name input.
With the help of special token set $sp$, We split the original description of each POI into $D_i =[x_i,y_i]$ for LLM. The text input $x_i$ and expected text output $y_i$ could be formulated as follows:
\begin{equation}
x_i =  [sp^{bos}_0,t_i^0,sp^{eos}_0,sp^{bos}_1,t_i^1,sp^{eos}_1]
\end{equation}
\begin{equation}
y_i =  [sp^{bos}_2,t_i^2,sp^{eos}_2,sp^{bos}_3,t_i^3,sp^{eos}_3...sp^{bos}_{n_k},t_i^{n_k},sp^{eos}_{n_k}]
\end{equation}
where $[...]$ means concatenation between different text.

We denote the LLM parameters as $\theta$ and employ a generative loss $L_{cp}$ for continual pretraining. 
\begin{equation}
L_{1} = - \frac{1}{\vert N_p \vert}\sum_{x_i,y_i \in N_p}^{\vert N_p \vert}\sum_{s=1}^{\vert y_i \vert} log(P_{\theta}({y_i}_s \vert x_i, {y_i}_{<s}))
\end{equation}
The input of LLM consists of restaurant's name and location $x_i$, aiming to predict other detail information $y_i$ such as POI's introduction, main dishes offered...
During the training process, the model learned the associations between similar dishes and built connections among geographical locations, POI names, menu and so on.

\subsection{Text Embedding via Contrastive Learning}

There is a huge gap between the normal decoder-only architecture of LLM and text embedding models, because language models train and infer in a way of autoregressive approach. To efficiently leverage the semantic information from the LLM, a critical problem we faced is how to transform the LM into a text embedding model. Common methods for converting a language model into one that could encode a sentence into a embedding include using the output embedding corresponding to some special tokens as the sentence embedding, or directly use some pooling methods on embeddings of all words in sentence to obtain the sentence embedding. Inspired by OpenAI \cite{neelakantan2022text}, we decide to adopt contrastive learning to convert the LLM into a text embedding model. In NLP field, contrastive learning methods usually construct positive sample pairs by applying corruptions such as dropout on text to generate positive sample pairs, random sampling for negative sample pairs. However, in RS, the importance of user, POI pairs in the food delivery context is very high, so we do not simply use corruption. Instead, we constructed positive and negative sample pairs from three perspectives: user-user, POI-POI, and user-POI.
\begin{figure*}[h]
  \centering
  \includegraphics[width=\linewidth]{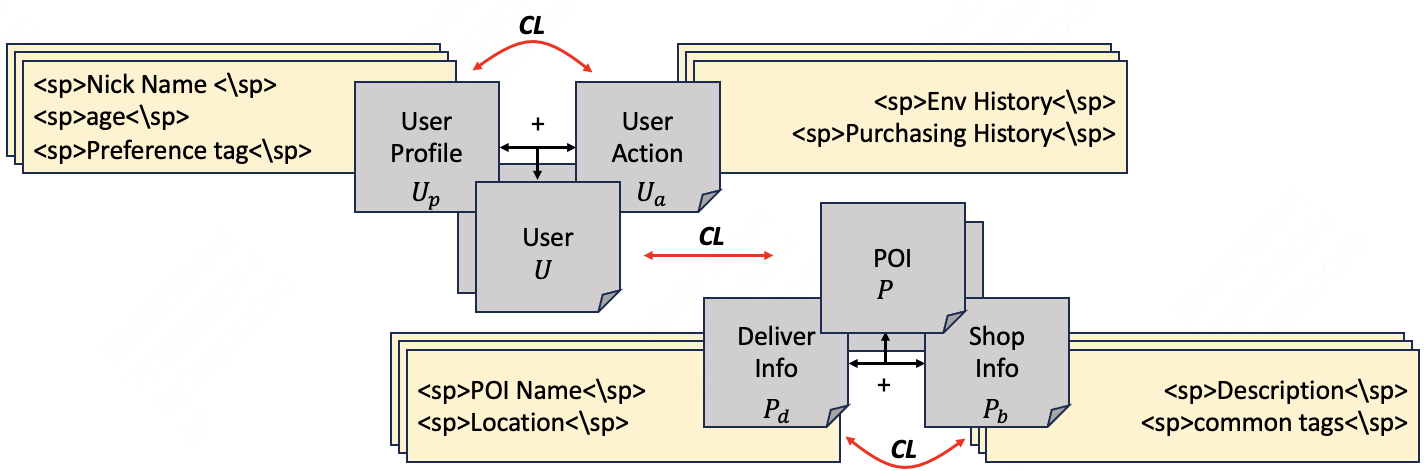}
  \caption{Positive pairs construction. Red curves represent alignment procedure.}
  \Description{samples fig 3}
  \label{fig:3}
\end{figure*}

As is shown in figure ~\ref{fig:3}, the LLM tends to in classify three kinds of input is positive or not in contrastive learning procedure. First of all, because of user-side text is added, we utilize a set special tokens $sp_u$ for user-side text similar to $sp_p$ in section 3.1. 
\begin{equation}
sp_{u} = \{ (sp^{bos}_k, sp^{eos}_k) \mid k \in \mathbb{N}, n_k+1 \leq k \leq n_k+n_q \}
\end{equation}
The $n_q$ represents the number of keywords in user-related text. We denote the union of two sets as $sp$.
\begin{equation}
sp = sp_p \cup sp_u
\end{equation}
Here, set $sp$ has $n_k+n_q$ elements.

We define a LLM-based score functiona $s_{\theta}$ for evaluating the similarity of inputs. The score functiona $s_{\theta}$ would calculate the similarity of two input text base on $P_{\theta}$. Similar to continual pretraining stage, the description $z$ of POI or user is divided into several segments, interspersed with special tokens in $SP$. Assuming that there's a series of text input $z=[sp^{bos}_0,z^0,sp^{eos}_0,sp^{bos}_1,z^1,sp^{eos}_1...sp^{bos}_l,z^l,sp^{eos}_l]$, it would be first tokenized into a series of tokens and mapping to a series of token embedding $\mathbf{Z}=[\mathbf{z_0},\mathbf{z_1}...\mathbf{z_m}]$, then sending to LLM after padding.
\begin{equation}
\mathbf{h}_{z}=MLP(LLM_{\theta}(z)_m)
\end{equation}
where $MLP$ is a linear layer to project the embedding of last token in last hidden layer to a continuous vector space for scoring.
\begin{equation}
s_{\theta}(z_1,z_2)=s(\mathbf{h}_{z_1}, \mathbf{h}_{z_2})
\end{equation}
Here, $s$ could be a similarity function such as cosine similarity; or negative form of the output of a distance function such as cosine distance. It could be symmetric or asymmetric. If $s$ is symmetric, then $s_{\theta}$ is symmetric; otherwise $s_{\theta}$ asymmetric.

\textbf{USER-USER} In user-user contrastive learning, we focus on user's profile text descriptions $U_p$ (such as nicknames, gender, etc.) and the user's action text descriptions $U_a$ (such as historical actions, environments, click and order price statistics...).

These two types of text descriptions are positive samples for the same user and negative samples for different users. During training, we adopt a contrastive learning approach, with the loss being the Info NCE Loss. For user-user contrastive loss $L^{UU}$, considering the general situation, we assume that $s_{\theta}$ is asymmetric. As a result, the form of contrastive loss is symmetrical and the sub loss $L^{UU}_{pa}$ could be fomulated as follows:
\begin{equation}
L^{UU}_{pa}=-\frac{1}{N_u}\sum_{i=1}^{N_u}\frac{exp(s_{\theta}(U_p, U_a+)/\tau)}{exp(s_{\theta}(U_p, U_a+)/\tau) + \sum exp(s_{\theta}(U_p,U_a-)/\tau)}
\end{equation}
\begin{equation}
L^{UU}_{ap}=-\frac{1}{N_u}\sum_{i=1}^{N_u}\frac{exp(s_{\theta}(U_a, U_p+)/\tau)}{exp(s_{\theta}(U_a, U_p+)/\tau) + \sum exp(s_{\theta}(U_a,U_p-)/\tau)}
\end{equation}
here $s_{\theta}$ is a score function for evaluating the similarity of inputs.symbol $+$ means corresponding positive sample and $-$ is negative samples that using in-batch negative sampling strategy.

The final loss of user-user contrastive loss is average of $U_p$ to $U_a$ and $U_a$ to $U_p$. 
\begin{equation}
L^{UU}= \frac{L^{UU}_{pa} + L^{UU}_{ap}}{2}
\end{equation}

\textbf{POI-POI} POI-POI contrastive learning considers the POI's food delivery business text description $P_d$ (such as name, location, menu) and the POI's basic information text description $P_b$ (such as shop normal introduction and some tags). Descriptions of the same POI are positive sample pairs, while those of different POIs form negative sample pairs. Similarly, the sub loss $L^{PP}_{db}$ POI-POI contrastive loss is as follows:
\begin{equation}
L^{PP}_{db}=-\frac{1}{N_p}\sum_{i=1}^{N_p}\frac{exp(s_{\theta}(P_d, P_b+)/\tau)}{exp(s_{\theta}(P_d, P_b+)/\tau) + \sum exp(s_{\theta}(P_d,P_b-)/\tau)}
\end{equation}

The POI-POI contrastive loss $L^{PP}$ is just like what $L^{uu}$ organizes:
\begin{equation}
L^{PP}= \frac{L^{PP}_{db} + L^{PP}_{bd}}{2}
\end{equation}
\textbf{USER-POI} The user-POI pairs are based on user's comprehensive text description $U$ (including user profile and user history texts): 
\begin{equation}
U =  [U_p,U_a]
\end{equation}

The POI's comprehensive text description $P$ (including food delivery business text and basic information text):
\begin{equation}
P =  [P_d,P_b]
\end{equation}

If a user paid for delivery in a POI, the comprehensive text descriptions of that user and POI constitute a positive sample pair; otherwise, they are a negative sample pair. 
\begin{equation}
L^{UP}_{up}=-\frac{1}{N_{seq}}\sum_{i=1}^{N_{seq}}\frac{exp(s_{\theta}(U, P+)/\tau)}{exp(s_{\theta}(U, P+)/\tau) + \sum exp(s_{\theta}(U,P-)/\tau)}
\end{equation}

To make it symmetric, $L^{UP}$ is arithmetic mean of $L^{UP}_{up}$ and $L^{UP}_{pu}$.
\begin{equation}
L^{UP}= \frac{L^{UP}_{up} + L^{UP}_{pu}}{2}
\end{equation}

Through the design of these three types of sample pairs, model can not only deeply understand the personalized characteristics of users' but also the delivery information of POIs' on user-user and POI-POI contrastive learning respectively. As $U_a$ records each delivery order and corresponding environment, LLM could precisely align user demands with POI offerings in complex and ever-changing environments.
\begin{equation}
L_{2}= \lambda_1*L^{UU} + \lambda_2*L^{PP} +\lambda_3*L^{UP}
\end{equation}
$\lambda$ here is for scaling the separate loss.

In inference, features that have appeared for users and POIs are sequentially fed into the LLM to infer and produce the corresponding description vectors from the feature level. The reason for not producing LLM vectors at the text level (aggregation of features) is due to the consideration that some features may change during the serving stage, such as weather, time slots, statistical features, etc. The cost of exhaustively generating all possible feature combination texts is exponential. Additionally, the user's long interaction sequence is also used to produce interaction sequence vectors.

\subsection{Information Alignment in RS}
In this section, contrastive learning is applied to alignment, maximizing the mutual information between the semantic information of the food delivery scene understood by LLM and the collaborative signals extracted from ID tokens, cross features, and statistical features by the RS model. The alignment not only extracts the shared information between semantic and collaborative signals but also reduces the noise from both, significantly enhancing the performance of the recommendation.

Specifically, our RS utilizes real-time scene embedding produced by the LLM after continual pretraining in stage 1 and fine-tuning in stage 2. Throughout this process, all parameters of the LLM are frozen, meaning the LLM acts like an encoder, converting a scene description into continuous vector representations to aid the training of the recommendation model. Since we want the model to be industry-friendly, the LLM should avoid participating in real-time inference as much as possible. We select $r$ real-time scene text features and decide to deal with scene text $s_i$ one by one, storing them in advance for the recommendation model to utilize, where  $0 \leq i \leq r$.
\begin{equation}
\mathbf{h}_{i}=LLM_\theta(s_i)_{sp^{eos}_i}
\end{equation}
Here $sp^{eos}_i$ represents the position of corresponding key real-time scene feature's end special token. We use the embedding of end special token in last hidden layer to represent real-time scene text. In reality, we selected 10 real-time scene text, that is, $r$ is set to 10.

We employ a bidirectional transformer encoder $\xi$ to tackle the problems that discrete real-time embeddings lack interaction. Since the LLM's outputs have already implicitly encoded the semantic associations between different scene features, so interactions provided by $\xi$ is necessary.

These semantic embeddings produced by LLM are stacked into a sequence and fed into $\xi$ for further aggregation and processing. To more effectively aggregate scene information, we introduced a trick at the beginning of the input sequence, a trainable aggregation token $<agg>$, corresponding to a vector $\mathbf{agg}$ that has the same dimension with real-time scene embedding. This special token plays a crucial role in aggregating all scene keyword and is trained within the transformer encoder to capture and integrate key information from different scene embedding. After processing of transformer encoder $\xi$, pooling method and a projection head (a normal MLP) is applied to the output embedding sequence, the result is denoted as $\mathbf{e}_{s}$, the real-time scene embedding with scene semantic information.
\begin{equation}
\mathbf{e}_{s} = MLP(pooling(\xi([\mathbf{agg},\mathbf{h}_{0},\mathbf{h}_{1}...\mathbf{h}_{r}])))
\end{equation}
Here $MLP$ projects the semantic embedding into the vector space of recommendation model and aligns the dimensions of pooling result and alignment objective. $r$ represents the number of real-time scene texts, as mentioned above.

Then we align $\mathbf{e}_{s}$ with target embedding $\mathbf{e}_t$ produced by recommendation model. $\mathbf{e}_t$ is the concatenation of user embedding and POI embedding in recommendation system. The procedure of alignment could be formulated as:
\begin{equation}
L^{st}_{cl}=-\frac{1}{N}\sum_{i=1}^{N}\frac{exp(s(\mathbf{e}_s, \mathbf{e}_t+)/\tau)}{exp(s(\mathbf{e}_s, \mathbf{e}_t+)/\tau) + \sum exp(s(\mathbf{e}_s,\mathbf{e}_t-)/\tau)}
\end{equation}
Here $s$ is the similarity scoring function. Considering the different symmetry properties of $s$, $N$ is batch size. total loss is the sum of two symmetric sub losses.
\begin{equation}
L_{cl}=\frac{L^{st}_{cl} + L^{ts}_{cl}}{2}
\end{equation}

This design not only improves the quality of both representations but also enables the model to more accurately grasp the connection between user needs and real-time scenes, leading to a significant improvement in recommendation results. 
With this approach, our recommendation system can provide more personalized and precise recommendations in real-time scenes.

During training, the loss is a weighted sum of the recommendation CTR loss $L_{ctr}$ and the contrastive learning loss $L_{cl}$. $L_{ctr}$ is BCE which is widely-used.
\begin{equation}
L_{3}= \beta_1*L_{ctr} + \beta_2*L_{cl}
\end{equation}

\section{Experiment}
\subsection{Experiment Setup}

\textbf{Dataset. } Our evaluation is conducted on a real-world dataset from Meituan Waimai. For the Click-Through Rate (CTR) prediction task, as shown in Table ~\ref{t:data}, we perform negative sampling on the log data collected from April 1st to April 7th, 2024, yielding approximately 3.5 billion training samples. For the test set, we uniformly sample the data from the subsequent day, April 8th, 2024, resulting in 84 million samples. Specifically, the input traditional features include user statistical features, real-time contextual features, user-POI cross-statistical features, POI statistical features, and the label, among others. Semantic features comprise descriptions of user foundational information, POI foundational information, as well as statistical descriptions of both user and POI.
\begin{table}[htb]
\caption{Statistics of the dataset.}
\label{t:data}
\begin{tabular}{cc}
\hline
Field                                                                         & Size         \\ \hline
\#Users                                                                       & 0.24 billion \\
\#PoIs                                                                       & 4.37 million \\
\#Records                                                                     & 3.5 billion   \\ \hline
\end{tabular}
\end{table}

During the continual pretraining phase, we select information from 4 million POIs to construct approximately 1B training corpora, which is then mixed with 9B Chinese corpora from Wudao \cite{yuan2021wudaocorpora}, forming a total of 10 billion pre-training textual corpora. In the fine-tuning phase, we extract 1 million samples from the historical behaviors of 2 million users for model fine-tuning. Additionally, all user and POI information data has been anonymized during the training of the LLM.

\begin{table*}[]
\caption{Performance comparison between our LARR and
baselines}
\label{p:data}
\begin{tabular}{l|ccccc}
\hline
 & Model & CTR AUC & CTCVR AUC & CTR GAUC & CTCVR GAUC \\ \hline
\multicolumn{1}{c|}{\multirow{2}{*}{Traditional Feature Learning Models}} & Wide\&Deep & 0.7912 & 0.8894 & 0.7011 & 0.6869 \\
\multicolumn{1}{c|}{} & DeepFM & 0.7918 & 0.8896 & 0.7009 & 0.6865 \\ \hline
\multirow{3}{*}{Context-Aware Recommendation Models} & P5 & 0.7439 & 0.8363 & 0.6422 & 0.6318 \\
 & PLE & 0.7983 & 0.8947 & 0.7088 & 0.6939 \\
 & LARR & \textbf{0.8030} & \textbf{0.8978} & \textbf{0.7107} & \textbf{0.6955} \\ \hline
\end{tabular}
\end{table*}

\textbf{Baseline. }To evaluate the performance of our experiments, we have selected a variety of classic experimental baselines, encompassing both traditional and semantic models. Specifically:
\begin{itemize}
\item {\textbf{Wide\&Deep} \cite{cheng2016wide}} is a machine learning model that integrates linear components (Wide) for handling sparse input features with deep neural networks (Deep) to learn feature interactions. It aims to leverage both memorization and generalization and is widely used in recommendation systems.
\item {\textbf{DeepFM} \cite{guodeepfm}} is a recommendation model that combines Factorization Machines (FM) with deep neural networks, designed to retain the advantages of FM while automatically learning high-order feature interactions.
\item {\textbf{P5} \cite{geng2022recommendation}} is a model based on the language model T5 that transforms the recommendation task into a text generation task. Additionally, given that P5 is a generative model, we fine-tuned it using the food delivery dataset to fulfill tasks such as CTR prediction.
\item {\textbf{PLE} \cite{tang2020progressive}} is a multi-task model based on the Mixed Model of Experts (MMOE) structure, which also serves as the baseline model in our model.
\item {\textbf{LARR}} is our proposed model, opts for the PLE model as the foundation for the traditional model component, and Baichuan2-7B \cite{baichuan2023baichuan2} for the semantic model part, with the aim of fully understanding the semantic information of real-time scenes and effectively integrating multimodal features to enhance recommendation results.
\end{itemize}

\textbf{Parameter Setting} To evaluate performance, we utilize Click-Through Rate (CTR AUC) and Conversion Rate (CTCVR AUC) as evaluation metrics, along with Group AUC (GAUC) for assessment.Our backbone LLM utilizes the open-sourced Baichuan2-7B \cite{baichuan2023baichuan2}. As detailed in Section 3, the backbone LLM first constructs a descriptive language corpus for all POIs on Meituan, upon which it performs continual pretraining. The epoch is set to 1, with a learning rate of 1e-4. The continual pretraining runs for approximately 96 hours on 32 A100-80G GPUs. Upon completion of continual pretraining, contrastive learning is conducted from three dimensions: "user-user," "POI-POI," and "user-POI" to construct the vector model, encoding scene features into 128-dimensional vectors and stacking them into sequences.  A 1-layer transformer encoder is employed to aggregate the encoded scene vectors, producing real-time scene information. During the alignment phase, an MLP is used as the projection head to project real-time scene information into the vector space of the alignment target. The projected vectors and alignment targets undergo contrastive learning with Info-NCE loss, treating vectors from the same sample as positive and those from different samples as negative. The temperature is set to 0.1, and the loss balancing coefficient is set to 0.05. The projected vectors and alignment targets, after concatenation, are used for CTR and CTCVR prediction. For fair comparison across all methods, we chose the following parameter configurations: a batch size of 2400, training for one epoch, with a learning rate of 8e-4. The tests run on 8 A100 GPUs, and for other parameters, we follow the best results from the original papers or source codes.


\subsection{Performance Comparison}
In this section, we provide a detailed comparison of the performance of our model, LARR, against various baselines. As shown in Table ~\ref{p:data}, an analysis of the experimental results reveals the following findings:

Firstly, we observe that traditional recommendation system models, such as Wide\&Deep and DeepFM, generally underperform more advanced deep learning models like PLE. This phenomenon suggests that complex deep network structures have significant advantages in learning feature interactions and capturing complex user interest patterns. By introducing deeper and more intricate interactions, these networks can more finely mine the underlying motivations behind user behavior, thereby achieving better accuracy in recommendations.

Secondly, our experimental results indicate that the semantic model P5 does not perform well in the food delivery recommendation task. This may suggest that in recommendation systems, traditional statistical features can sometimes more directly reflect the actual interests of users compared to semantic features. Although semantic models can provide rich contextual information, statistical analysis of user historical behavior may more directly and effectively reveal user preferences in practical recommendation scenes.

Finally, our model, LARR, outperforms all baseline models. Compared to the PLE model, LARR achieves a significant 0.58\% improvement in CTR AUC and also gains an 0.34\% increase in CTCVR AUC. These results not only confirm the superiority of our model but also emphasize the effectiveness of LARR in integrating the semantic information of real-time scenes with recommendation modality information. Through this integration, LARR can capture users' immediate needs and potential interests more accurately, significantly enhancing the overall performance and user satisfaction of the recommendation.

\subsection{Ablation Study}
We introduce a novel multimodal information fusion framework, LARR. Initially, based on Food Delivery recommendation data, we perform continual pretraining and fine-tuning on LLMs to infuse domain knowledge and enhance the LLMs' understanding of real-time scenes.
Subsequently, in the CTR task, we effectively integrate the real-time semantic information with recommendation modality information, thereby improving recommendation results. To verify the effectiveness of these designs, we conducted ablation studies.
\begin{itemize}
\item {\textbf{LARR-PLE} Variant}: The model variant that completely removes the LLM module, essentially the original PLE model.

\item {\textbf{LARR-noCPT} Variant}: The model variant that omits the continual pretraining phase of the LLM.

\item {\textbf{LARR-noFT} Variant}: The model variant that excludes the vector fine-tuning phase of the LLM.

\item {\textbf{LARR-noAL} Variant}: The model variant that removes the multimodal feature alignment stage, relying solely on a simple feature concatenation strategy.

\item {\textbf{LARR} Complete}: The complete LARR framework.
\end{itemize}
\begin{table}[]
\caption{Performance comparison of LARR and different
variants.}
\label{p:data2}
\begin{tabular}{ccc}
\hline
Model          & CTR AUC         & CTCVR AUC       \\ \hline
LARR-PLE       & 0.7983          & 0.8947          \\
LARR-noCPT  & 0.8017          & 0.8969          \\
LARR-noFT  & 0.7985          & 0.8948          \\
LARR-noAL & 0.8007          & 0.8961          \\
LARR-Complete  & \textbf{0.8030} & \textbf{0.8978} \\ \hline
\end{tabular}
\end{table}
The results are illustrated in the figure ~\ref{p:data2}.
The improvement of LARR Complete over LARR-noCPT Variant indicates that domain-specific continual pretraining of LLMs can significantly enhance the model's understanding of domain knowledge, thus capturing the semantic information of user and POI more accurately. The comparison between LARR Complete and LARR-noFT Variant demonstrates that the base model of LLM, without vector fine-tuning, does not yield ideal results; however, an LLM that has undergone vector fine-tuning can more effectively understand the personalized features of users and POIs, generating more expressive feature vectors. The comparison between LARR Complete and LARR-noAL Variant underscores the importance of multimodal feature alignment. Simple feature concatenation cannot compensate for the spatial differences between semantic features and traditional collaborative features, whereas a carefully designed multimodal alignment mechanism can achieve effective integration of features from different modalities. Ultimately, LARR Complete surpasses all variants on all metrics, validating the effectiveness of our adopted strategies in enhancing recommendation quality. Overall, the success of LARR is attributed to its profound understanding of complex user behaviors and diverse POI features in real-time scenes, as well as its efficient capability to fuse multimodal information. Our research offers fresh perspectives on how to leverage large language models and contrastive learning to enhance recommendation results.

\subsection{Hyperparameter Analysis}
\begin{figure}[ht]
  \centering
  \begin{subfigure}[b]{0.49\linewidth}
    \includegraphics[width=\linewidth]{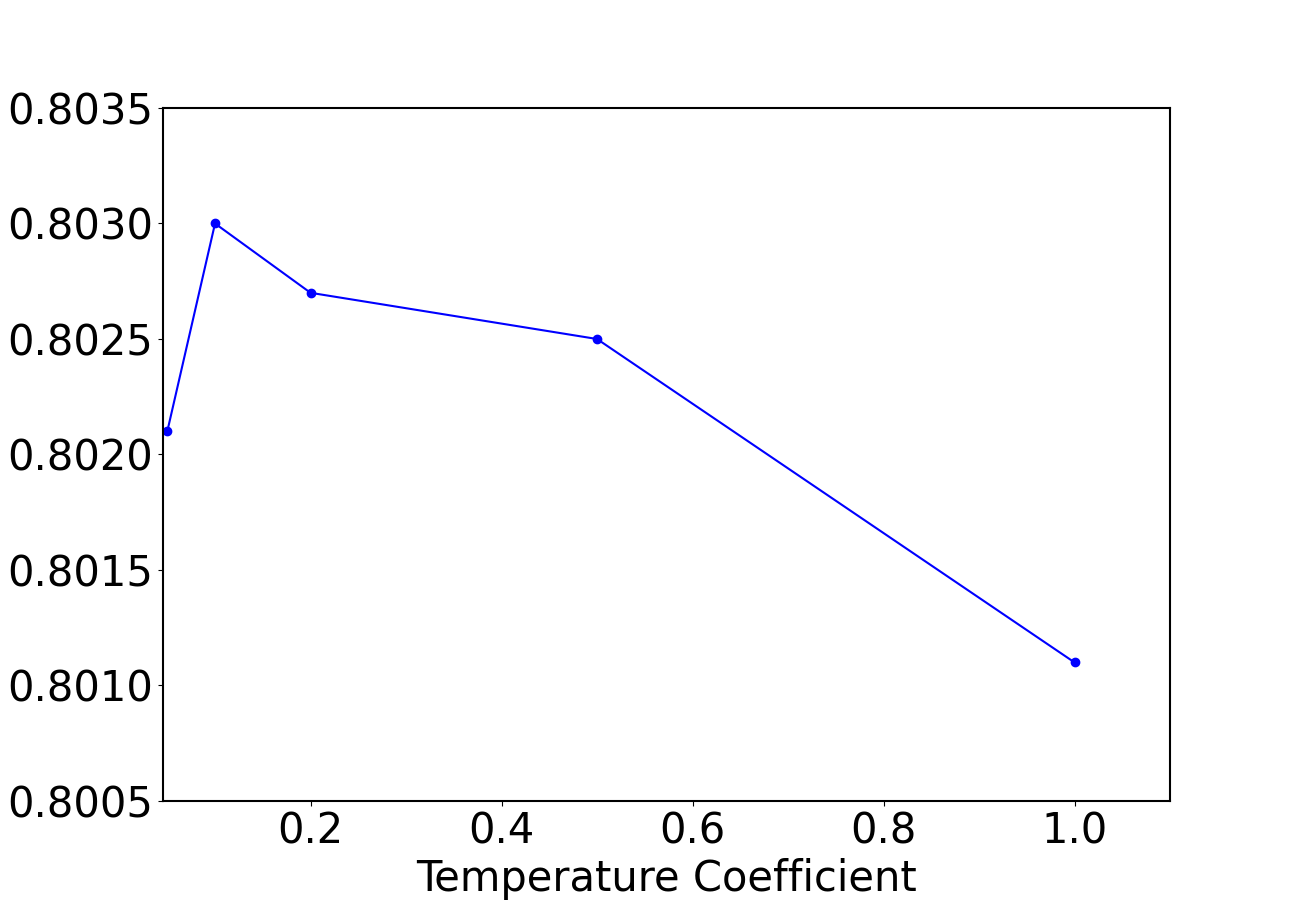}
    \caption{Temperature Coefficient vs CTR AUC}
    \label{fig:temp}
  \end{subfigure}
  \hfill 
  \begin{subfigure}[b]{0.49\linewidth}
    \includegraphics[width=\linewidth]{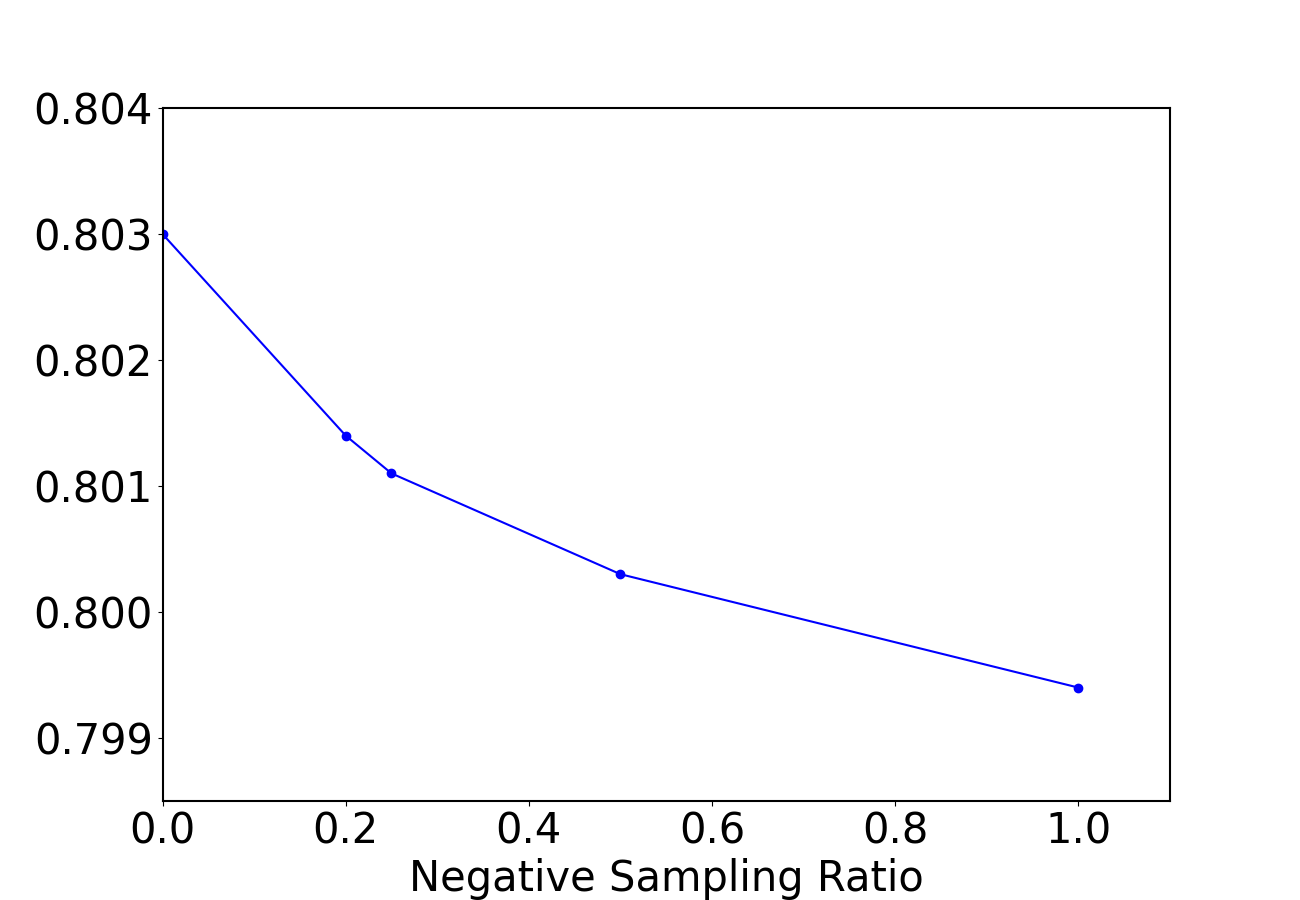}
    \caption{Negative Sampling Ratio vs CTR AUC}
    \label{fig:negsamp}
  \end{subfigure}
  \caption{Hyperparameter Analysis}
  \label{fig:hyperparameters}
\end{figure}

In Section 4.4, our aim is to align the deep semantic information of food delivery scenes as understood by LLMs with the collaborative signals extracted from ID tokens, cross-features, and statistical features by the conventional fine-tuning models. We seek to extract the shared information between semantic and collaborative data to enhance model performance. During the alignment phase of the model, we focus on two key hyperparameters: the temperature coefficient and the negative sample sampling ratio. The temperature coefficient plays a crucial role in the loss function of contrastive learning, adjusting the sensitivity of the loss and influencing the model's ability to distinguish between positive and negative samples. A lower temperature coefficient increases the difficulty for the model to differentiate samples, causing it to focus more on challenging sample pairs. Conversely, the negative sample sampling ratio controls the ratio of negative to positive samples during the contrastive learning process. In this section, we conduct experimental analysis on these two hyperparameters with respect to CTR AUC, with results depicted in Figure ~\ref{fig:hyperparameters}

Figure ~\ref{fig:temp} illustrates the relationship between the temperature coefficient and experimental performance. We find that when the temperature coefficient is set to 0.1, the model achieves optimal performance, indicating that at this specific setting, the model can effectively learn from the data and improve its discrimination ability. However, if the temperature coefficient is set too low or too high, the learning efficiency of the model is impacted, leading to decreased performance. On the other hand, the experimental results regarding the negative sample sampling ratio, as shown in Figure ~\ref{fig:negsamp}, indicate that introducing negative sample sampling in our setup does not lead to the anticipated performance improvement. Instead, we observe a decline in model performance after employing negative sample sampling. This may be due to the fact that, in certain data and tasks, increasing the number of negative samples does not provide additional information but may instead introduce noise or cause the model to overly focus on negative samples, thereby affecting the model's learning efficiency with positive samples.

\subsection{Case Study}
We present a case study to further demonstrate the model's proficiency. Given that out-of-area ordering (where users place orders from locations other than their usual residence for various reasons) has always been an important scenario in food delivery services, we are particularly interested in whether the model can generate good recommendations for out-of-area users based on relevant semantics. We selected some random users who have never been to Guangdong, with the scene of "user visiting Guangdong for vacation", designed a simple recall experiment to observe the details of the top 10 POIs in terms of recall scores. A typical result is as follows, with each response presented as location; main dish. Sensitive information such as POI names and details have been omitted.

Top1: Guangzhou; Rice Noodle Roll

Top2:. Chaoshan; Stir-fried Beef Rice Noodles

Top3:. Guangzhou; Stir-fried Beef Rice Noodles

Top4:. Shenzhen; Seafood Claypot Congee

Top5: Guangzhou; Cantonese Barbecue

Top6: Guangzhou; Claypot Rice

Top7: Guangzhou; Rice Noodle Roll

Top8: Guangzhou; Sweet Soup

Top9: Dongguan; Sweet Soup

Top10: Foshan; Claypot Rice

In this case, Guangdong appeared the most frequently (5 times), with recalled POI primarily offering dishes such as Rice Noodle Roll, Stir-fried Beef Rice Noodles, Cantonese Barbecue, Claypot Rice, and Sweet Soup. ChaoShan, Shenzhen, Dongguan, and Foshan each appeared once or twice. This geographical distribution reflects the culinary diversity and popularity of Guangzhou as the capital city of Guangdong Province, which aligns with our survey findings that these dishes are representative of Guangdong's regional specialties. 

The recall results cover a variety of classic Cantonese dishes, indicating that the model performs well in capturing local characteristics. The recall results show that the model can identify multiple different types of dishes and accurately match them to the provided scene text, this diversity is positive for enhancing user satisfaction and meeting a wider range of user needs.

\subsection{Online A/B Test}
LARR demonstrated its practical value in an online A/B test within the Meituan recommendation system. During the testing period from April 15th to April 21st, 2024, LARR achieved a 2.5\% increase in CTR and a 1.2\% growth in GMV (Gross Merchandise Volume) compared to the current baseline model. These results confirm the effectiveness of LARR in understanding user needs and enhancing user experience. The increase in CTR indicates a higher acceptance of recommended content by users, which may be attributed to LARR's more accurate capture of user interests, thus providing recommendations that better align with user expectations. The growth in GMV directly reflects the positive impact of LARR on conversion rates and business revenue. This online test verified the practicality and effectiveness of LARR within the Meituan recommendation system.

\section{Conclusion}
To tackle the lack of understanding of semantic information in RS and the efficiency challenges in LLM-based CTR models, this paper proposes a novel LARR method. LARR continues to pretrain and fine-tune the LLM on some corpora with an enlarged vocabulary, using contrastive learning to align semantic signals and collaborative signals, leverage shared information, and reduce noise. Extensive online and offline experimental results demonstrate that LARR achieves low latency and enhances CTR performance, offering fresh insights for the practical deployment of LLM-based CTR models.

\bibliographystyle{ACM-Reference-Format}
\bibliography{ref}

\appendix

\end{document}